\RequirePackage{fix-cm}
\documentclass[smallextended]{svjour3}       
\smartqed  
\usepackage{graphicx}
\def\be{\begin{equation}}
\def\ee{\end{equation}}
\def\bea{\begin{eqnarray}}
\def\eea{\end{eqnarray}} 
\def\nnb{\nonumber}
\begin{document}

\title{
ANCs of the bound states of $^{16}$O
deduced from elastic $\alpha$-$^{12}$C scattering data
}


\author{
Shung-Ichi Ando
}


\institute{S.-I. Ando \at
              Department of Display and Semiconductor Engineering, 
Sunmoon University, Asan, Chungnam 31460, Republic of Korea \\
              \email{sando@sunmoon.ac.kr}           
}

\date{Received: date / Accepted: date}

\maketitle

\begin{abstract}
Asymptotic normalization coefficients (ANCs) of the $0_1^+$, $0_2^+$,
$1_1^-$, $2_1^+$, $3_1^-$ ($l_{i th}^\pi$) bound states of 
$^{16}$O are deduced 
from the phase shift data of elastic $\alpha$-$^{12}$C scattering 
at low energies. 
$S$ matrices of elastic $\alpha$-$^{12}$C scattering are constructed 
within cluster effective field theory (EFT), in which both bound and 
resonant states of $^{16}$O are considered. 
Parameters in the $S$ matrices are fitted to the precise phase shift 
data below the $p$-$^{15}$N breakup energy for the partial waves 
of $l=0,1,2,3,4,5,6$, and  
the ANCs are calculated by using the wave function normalization factors
of $^{16}$O propagators for $l=0,1,2,3$. 
We review the values of ANCs, 
which are compared with other results in the literature, and discuss
uncertainties of the ANCs obtained from the elastic $\alpha$-$^{12}$C 
scattering data in cluster EFT. 

\end{abstract}

\section{Introduction}
\label{intro}

Radiative $\alpha$ capture on $^{12}$C, $^{12}$C($\alpha$,$\gamma$)$^{16}$O,
is a fundamental reaction of nuclear astrophysics, which determines
the C/O ratio during helium burning processes in a star. 
The radiative capture rate of  $^{12}$C($\alpha$,$\gamma$)$^{16}$O
at Gamow-peak energy, 
which represents an energy at the typical temperature 
of helium burning processes, $E_G=0.3$~MeV,
has not been measured in an experimental facility 
because of the Coulomb barrier. 
One needs to employ a theoretical model to extrapolate the reaction rate
to $E_G$, where the model parameters are fixed 
(or the model is tested) by using the experimental data measured 
at a few MeV energy. 
During the last half-century, many experimental and theoretical studies were 
devoted to the reaction. 
One may refer to, e.g., 
Refs.~\cite{f-rmp84,bb-npa06,detal-rmp17,sa-epja21} 
for review.

Effective field theory (EFT) is a new theoretical method for the study
of $^{12}$C($\alpha$,$\gamma$)$^{16}$O reaction. 
While EFT is fairly popular 
for the study of few-body systems in hadron and nuclear physics;
it provides a systematic method to calculate a reaction amplitude at 
low energy involving a perturbative expansion scheme in terms of $Q/\Lambda_H$
where $Q$ denotes a typical momentum scale of a reaction and 
$\Lambda_H$ does a high momentum scale whose degrees of freedom are 
integrated out of the effective Lagrangian~\cite{hkvk-rmp20}. 
The theory reproduces the effective range expansion~\cite{b-pr49} 
and, moreover, can include higher-order corrections as well as 
external currents which are devices to incorporate electromagnetic 
and weak interactions in the 
theory~\cite{am-plb98,af-prd07}.
Recently, it was reported that EFT calculations and the zero channel 
radius limit of Wigner's $R$-matrix theory lead to the identical results
for $np$ scattering in $^1S_0$ channel and $^3$H($d$,$n$)$^{4}$He 
reaction~\cite{hbp-prc14}. 

EFT has been employed in the studies of many reactions, which 
are crucial for nuclear astrophysics, 
e.g., neutron $\beta$-decay~\cite{aetal-plb04},
radiative neutron capture on a proton 
at big-bang nucleosynthesis energy~\cite{cs-prc99,r-npa00,achh-prc06},
proton-proton fusion in hydrogen burning 
processes~\cite{kr-npa99,bc-plb01,aetal-plb08}, 
solar-neutrino reactions on the deuteron~\cite{bck-prc01,ash-prc20}, 
and radiative proton capture on $^{15}$N in the CNO cycle~\cite{sao-prc22}. 
During the last decade, we have been studying the construction of EFT 
for the $^{12}$C($\alpha$,$\gamma$)$^{16}$O reaction and applied the 
formalism to the studies of 
various cases of elastic $\alpha$-$^{12}$C scattering 
at low energy~\cite{sa-epja16,sa-prc18,sa-jkps18,sa-prc22}, 
$E1$ transition of $^{12}$C($\alpha$,$\gamma$)$^{16}$O 
reaction~\cite{sa-prc19}, 
and $\beta$-delayed $\alpha$ emission from $^{16}$N~\cite{sa-epja21}.
In the present contribution, we discuss the calculation of 
asymptotic normalization coefficients (ANCs) of $0_1^+$, $0_2^+$,
$1_1^-$, $2_1^+$, $3_1^-$ ($l^\pi_{ith}$) bound states of $^{16}$O as two-body
$\alpha$-$^{12}$C cluster system, deduced from the phase shift 
data of elastic $\alpha$-$^{12}$C scattering 
in cluster EFT~\cite{sa-prc23,sa-23}. 

The use of an ANC is the standard
method in nuclear astrophysics to estimate a nuclear reaction rate
at low energy; it determines the overall strength of the reaction rate
within potential model and $R$-matrix analysis~\cite{tn-09,d-03}. 
In cluster EFT, 
an ANC of a bound state, as a two-body cluster system, of a nucleus
can be obtained by using the wave function normalization factor 
of the propagator of the nucleus. 
The propagators are parameterized by means of the effective range 
expansion and those parameters are fixed by using
elastic scattering data of the two-body system of the nucleus. 
Thus, the ANCs of bound states of a nucleus can be straightforwardly
calculated within the framework of cluster EFT. 
In addition, we construct 
the $S$ matrices of elastic scattering to include the bound and resonant
states of $^{16}$O~\cite{sa-prc23}. 
In the present paper, we first review the formalism
to construct the $S$ matrices of elastic $\alpha$-$^{12}$C scattering in 
cluster EFT and, 
after obtaining the ANCs of 
$0_1^+$, $0_2^+$, $1_1^-$, $2_1^+$, $3_1^-$ bound states of $^{16}$O 
by fitting 
the effective range parameters to the accurate phase shift data,
we discuss the issues of ANCs deduced from the phase shift data 
comparing with other results in the literature.

The present work is organized as follows.
In Sec. 2, the formalism of $S$ matrices of elastic $\alpha$-$^{12}$C 
scattering at low energy in cluster EFT is reviewed and the parameters 
of the $S$ matrices
are fitted to the phase shift data. In Sec. 3, the result of the ANCs 
of bound states of $^{16}$O are displayed and discussed comparing with 
other results in the literature. 
Finally, in Sec. 4, the summary of this work is presented. 

\section{$S$ matrices of elastic $\alpha$-$^{12}$C scattering}
\label{sec:1}

The $S$ matrices of elastic $\alpha$-$^{12}$C scattering 
for $l$-th partial wave states are given in terms of 
phase shifts, $\delta_l$, and elastic scattering amplitudes, $\tilde{A}_l$, as 
\bea
S_l = e^{2i\delta_l} 
= 1 + 2ip\tilde{A}_l\,.
\label{eq;S_l}
\eea
We now assume that the phase shifts can be decomposed as, e.g.,
\bea
\delta_l = \delta_l^{(bs)} + \delta_l^{(rs1)}
+ \delta_l^{(rs2)}
\,,
\eea
where $\delta_l^{(bs)}$ is a phase shift generated from 
a bound state, and $\delta_2^{(rsN)}$ with $N=1,2$
are those from the first and second resonant states, and 
each of those phase shifts may have
a relation to a corresponding scattering amplitude as
\bea
e^{2i\delta_l^{(ch)}} &=& 1 + 2ip\tilde{A}_l^{(ch)}\,,
\eea
where $ch(annel) = bs, rs1, rs2$,
and $\tilde{A}_l^{(bs)}$ and $\tilde{A}_l^{(rsN)}$ with $N=1,2$
are the amplitudes for the binding part and the first and second resonant
parts of the amplitudes,
which are derived from the effective Lagrangian in Ref.~\cite{sa-prc23}.
The total amplitudes $\tilde{A}_l$ for the nuclear reaction part
in terms of the three amplitudes,
$\tilde{A}_l^{(bs)}$ and $\tilde{A}_l^{(rsN)}$
with $N=1,2$, read
\bea
\tilde{A}_l &=&
\tilde{A}_l^{(bs)}
+ e^{2i\delta_l^{(bs)}} \tilde{A}_l^{(rs1)}
+ e^{2i(\delta_l^{(bs)}+\delta_l^{(rs1)})} \tilde{A}_l^{(rs2)}
\,.
\label{eq;tldAl}
\eea
We note that the total amplitudes 
are not obtained as
the summation of the amplitudes, 
but the additional phase factors appear on them. 

\begin{figure}[t]
  \includegraphics[width=0.9\textwidth]{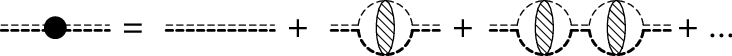}
\caption{
Diagrams for dressed $^{16}$O propagators.
A thick and thin double dashed line with and without a filled circle
represent dressed and bare $^{16}$O propagator, respectively. 
A thick (thin) dashed line represents a propagator of $^{12}$C ($\alpha$),
and a shaded blob in loop diagrams does the Coulomb green's function. 
}
\label{fig:16O_propagator}       
\end{figure}
\begin{figure}[t]
  \includegraphics[width=0.21\textwidth]{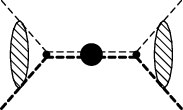}
\caption{
Diagram for elastic $\alpha$-$^{12}$C scattering amplitudes. 
A shaded blob represents an initial or final Coulomb wave function,
and a thick and thin double-dashed line with a filled circle a 
dressed $^{16}$O propagator. See the caption of Fig.~\ref{fig:16O_propagator}
as well. 
}
\label{fig:Amplitude}       
\end{figure}

The amplitudes are calculated by using the diagrams in 
Figs.~\ref{fig:16O_propagator} and \ref{fig:Amplitude}.
For the bound state amplitudes, $\tilde{A}_l^{(bs)}$ with $l=0,1,2,3$,
one has
\bea
\tilde{A}^{(bs)}_l &=&
\frac{C_\eta^2 W_l(p)}{
K_l(p)
-2\kappa H_l(p)
}
\label{eq;Aer2}
\label{eq;A2_nr}
\,,
\eea
where $C_\eta^2 W_l(p)$ in the numerator of the amplitude
is calculated from the initial and final state Coulomb interactions
in Fig.~\ref{fig:Amplitude}; $p$ is the magnitude of
relative momentum of the $\alpha$-$^{12}$C system in the center of mass
frame, $p=\sqrt{2\mu E}$: $\mu$ is the reduced mass of $\alpha$ and $^{12}$C,
and 
\bea
W_l(p) = \left(\frac{\kappa^2}{l^2} + p^2\right) W_{l-1}(p)\,,
\ \  W_0(p) = 1\,,
\ \ C_\eta^2 = \frac{2\pi\eta}{\exp(2\pi\eta)-1}\,,
\eea
where $\eta$ is the Sommerfeld parameter, $\eta=\kappa/p$:
$\kappa$ is the inverse of the Bohr radius,
$\kappa = Z_\alpha Z_{12C} \alpha_E\mu$,
where $Z_A$ is the number of protons
in a nucleus, and $\alpha_E$ is the fine structure constant.
The function $-2\kappa H_l(p)$ in the denominator of the amplitude
is the Coulomb self-energy term which is calculated
from the loop diagram in Fig.~\ref{fig:16O_propagator}, and one has
\bea
H_l(p) &=& W_l(p) H(\eta)\, ,
\ \ \
H(\eta) = \psi(i\eta) + \frac{1}{2i\eta} -\ln(i\eta)\,,
\eea
where $\psi(z)$ is the digamma function.
The nuclear interaction is represented
in terms of the effective range parameters
in the function $K_l(p)$ in
the denominator of the amplitude in Eq.~(\ref{eq;Aer2}).
As discussed in Ref.~\cite{sa-prc18}, large and significant contributions
to the series of effective range expansions,
compared to that evaluated from the phase shift data at the lowest energy
of the data, $E_\alpha=2.6$~MeV,
appear from the Coulomb self-energy term, $-2\kappa H_l(p)$;
$E_\alpha$ is the $\alpha$ energy in the laboratory frame. 
To subtract those contributions,
we include the effective range terms up to
$p^6$ order for $l=0,1,2$ and those up to $p^8$ order for $l=3$
as counter terms.
Thus, we have
\bea
K_l(p) &=&
-\frac{1}{a_l}
+\frac12 r_lp^2
-\frac14 P_l p^4
+Q_l p^6
-R_l p^8\,,
\eea
where $a_l$, $r_l$, $P_l$, $Q_l$, $R_l$ are effective range parameters:
note that $R_l=0$ for $l=0,1,2$.

We fix a parameter or two among the four (or five for $l=3$) 
effective range parameters,
$a_l$, $r_l$, $P_l$, $Q_l$ (and $R_3$) by using the condition that
the inverse of the scattering amplitude $\tilde{A}_l^{(bs)}$ vanishes at
the binding energy of a bound state of $^{16}$O.
Thus, the denominator of the scattering amplitude,
\bea
D_l(p) = K_l(p) -2\kappa H_l(p) \,,
\label{eq;binding_pole}
\eea
vanishes at $p=i\gamma_{l(i)}$ where $\gamma_{l(i)}$ are the binding momenta
of the $0_1^+$, $0_2^+$, $1_1^-$, $2_1^+$, $3_1^-$ ($l^\pi_{i-th}$) states 
of $^{16}$O;
$\gamma_{l(i)} = \sqrt{2\mu B_{l(i)}}$ where $B_{l(i)}$
are the binding energies of the bound states of $^{16}$O
associated with the $\alpha$-$^{12}$C breakup threshold energy.
At the binding energies, one has the wave function normalization factors
$\sqrt{Z_{l(i)}}$ for the bound states of $^{16}$O in the dressed $^{16}$O
propagators as
\bea
\frac{1}{D_l(p)} = \frac{Z_{l(i)}}{E+B_{l(i)}} + \cdots\,,
\eea
where the dots denote the finite terms at $E=-B_{l(i)}$, and 
one has
\bea
\sqrt{Z_{l(i)}} = \left(
\left|
\frac{dD_l(p)}{dE}
\right|_{E=-B_{l(i)}}
\right)^{-1/2}=
\left(
2\mu
\left|
\frac{dD_l(p)}{dp^2}
\right|_{p^2=-\gamma_{l(i)}^2}
\right)^{-1/2}\,.
\label{eq;sqrtZl}
\eea
The wave function normalization factor $\sqrt{Z_{l(i)}}$ is multiplied to a
reaction amplitude when a bound state appears in the initial or 
final state of a reaction.

the ANCs $|C_b|_{l(i)}$ for the bound states of $^{16}$O are calculated 
by using the formula of Iwinski, Rosenberg, and Spruch~\cite{irs-prc84}
\bea
|C_b|_{l(i)} = \frac{\gamma_{l(i)}^l}{l!} \Gamma(l+1+\kappa/\gamma_{l(i)})
\left( \left| \frac{dD_l(p)}{dp^2}\right|_{p^2 = - \gamma_{l(i)}^2}
\right)^{-1/2} \ \,, 
\label{eq;Cb}
\eea
where $\Gamma(x)$ is the gamma function, and one may notice that the ANCs
are proportional to the wave function normalization factor $\sqrt{Z_{l(i)}}$
comparing Eqs.~(\ref{eq;sqrtZl}) and (\ref{eq;Cb}).

The amplitudes for the resonant states may be obtained
in the Breit-Wigner-like expression as
\bea
\tilde{A}_l^{(rsN)} &=&
-
\frac{1}{p}
\frac{\frac12\Gamma_{(li)}(E) }{E-E_{R(li)}
+ R_{(li)}(E) + i\frac12\Gamma_{(li)}(E)}\,,
\label{eq;A2_rsN}
\eea
with
\bea
\Gamma_{(li)}(E) &=& \Gamma_{R(li)}
\frac{pC_\eta^2W_l(p)}
     {p_rC_{\eta_r}^2W_l(p_r)}\,,
\\
R_{(li)}(E) &=& a_{(li)}(E-E_{R(li)})^2 + b_{(li)}(E-E_{R(li)})^3
\,,
\label{eq;R}
\eea
where $E_{R(li)}$ and $\Gamma_{R(li)}$ are the energy and width of 
resonant $l_{i-th}$ states, and $p_r$ and $\eta_r=\kappa/p_r$ are 
the momenta and Sommerfeld factors at the resonant energies:
we suppressed the $i$ index for them. 
The functions $R_{(li)}(E)$ have
the second and third order corrections expanded around $E=E_{R(li)}$,    
where the coefficients, $a_{(li)}$ and $b_{(li)}$, are fitted 
to the shapes of resonant peaks. 

Using the relations for the amplitudes
in Eqs.~(\ref{eq;A2_nr}) and (\ref{eq;A2_rsN}),
the $S$ matrices in Eq.~(\ref{eq;S_l}) are obtained 
in a simple and transparent expression as
\bea
e^{2i\delta_l} &=&
\frac{K_l(p) - 2\kappa Re H_l(p) + ipC_\eta^2W_l(p)}
     {K_l(p) - 2\kappa Re H_l(p) - ipC_\eta^2W_l(p)}
\nnb \\ && \times
\prod_i
\frac{E - E_{R(li)} + R_{(li)}(E) - i\frac12\Gamma_{(li)}(E)}
     {E - E_{R(li)} + R_{(li)}(E) + i\frac12\Gamma_{(li)}(E)}
\,.
\label{eq;exp2idel_l}
\eea

\begin{table}
\begin{center}
\caption{
Bound and resonant ($l^\pi_{i-th}$) states of $^{16}$O,
which are considered to construct the $S$ matrices
for $l$-th partial wave states of elastic $\alpha$-$^{12}$C scattering.
The bound states are presented in the second column where we also introduced
a low energy background contribution for $l=6$.
The resonant states in the third column appear in the energy range of the 
experimental data,
$2.6~\textrm{MeV} \le E_\alpha \le 6.62$~MeV, and those in the fourth column
do at $6.62~\textrm{MeV} < E_\alpha$.
}
\label{table;states}
\begin{tabular}{c | c c c}
$l$ & (Bound states) & $2.6~\textrm{MeV} \le E_\alpha \le 6.62$~MeV, &
6.62~MeV $< E_\alpha$ \cr \hline
0 & $0_1^+, 0_2^+$ & $0_3^+$ & $0_4^+$ \cr
1 & $1_1^-$ & $1_2^-$ & $1_3^-$ \cr
2 & $2_1^+$ & $2_2^+$, $2_3^+$  & $2_4^+$ \cr
3 & $3_1^-$ & $3_2^-$ & $3_3^-$ \cr
4 & --      & $4_1^+$, $4_2^+$ & $4_3^+$ \cr
5 & --      & --      & $5_1^-$ \cr
6 & (bg)    &  --     & $6_1^+$ \cr 
\hline
\end{tabular}
\end{center}
\end{table}
In Table \ref{table;states},
we display a list of the bound and resonant states of $^{16}$O,
which are included in the $S$ matrices. 
Each of the amplitudes
has four parameters in general (five effective range parameters for the 
$3_1^-$ state) while $0_1^+$ and $0_2^+$ bound states are included 
in the single set of effective range parameters and the resonant $1_2^-$ and
$3_2^-$ states can be described in terms of the effective range parameters
for the $1_1^-$ and $3_1^-$ states, respectively. 
One or two effective range parameters are fixed by using the binding 
energies.
(We also include an amplitude for $l=6$, 
as a low-energy background contribution, 
and two effective range parameters, $r_6$ and $P_6$,
are retained.)  
For the resonant states appearing in the experimental data,
$a_{(li)}$ and $b_{(li)}$ parameters are suppressed for the very sharp 
$0_3^+$, $2_2^+$, $4_2^+$ resonant states. 
For the resonant states at $E_\alpha > 6.62~\textrm{MeV}$, 
representing the background
contributions from high energy, 
the resonant energies and widths are fixed 
by using the experimental data. Thus, we have 6, 5, 9, 6, 8, 2, 4 parameters 
to fix the experimental phase shift data for $l=0,1,2,3,4,5,6$, respectively.  

\begin{figure}[t]
  \includegraphics[width=0.5\textwidth]{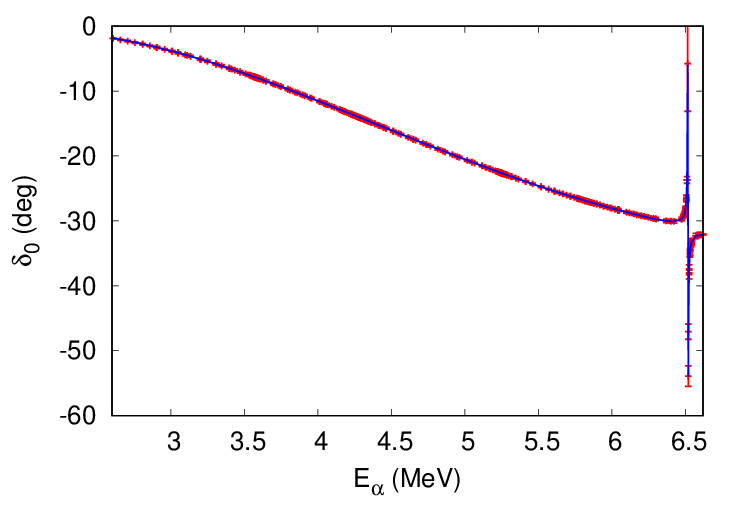}
  \includegraphics[width=0.5\textwidth]{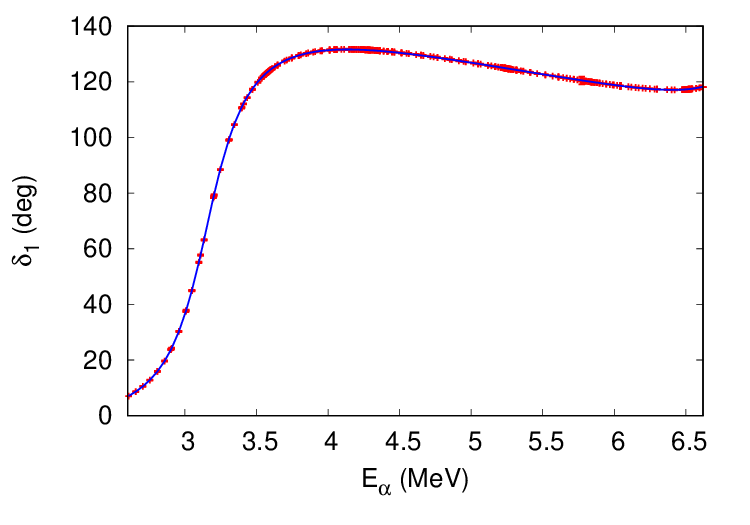}
  \includegraphics[width=0.5\textwidth]{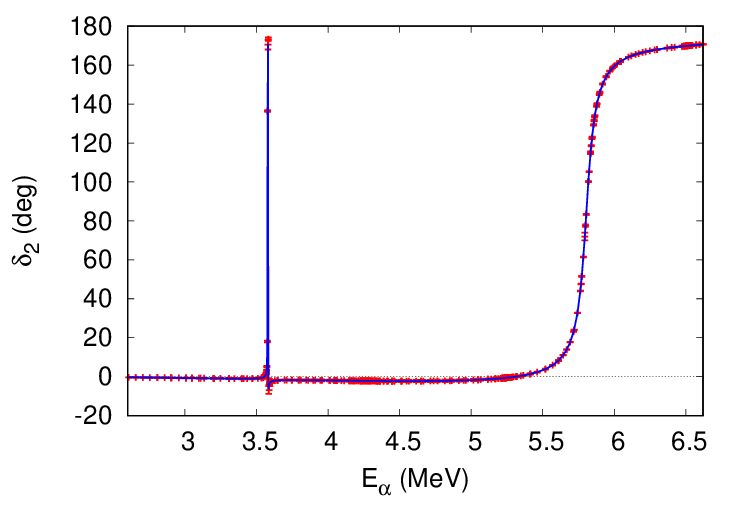}
  \includegraphics[width=0.5\textwidth]{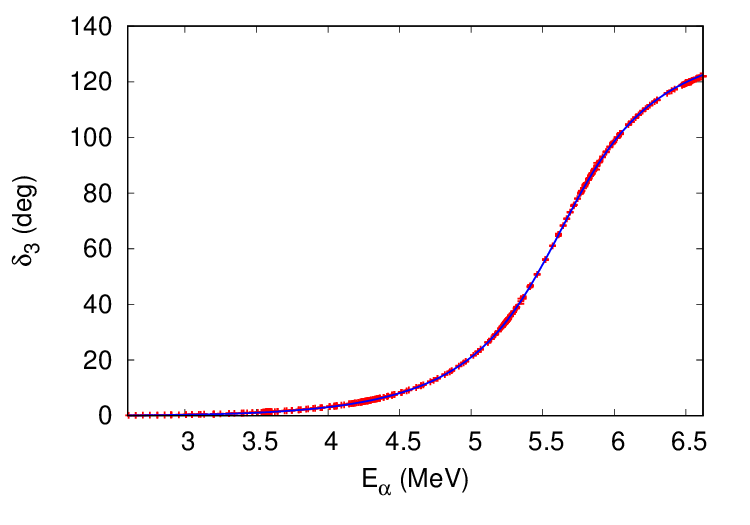}
  \includegraphics[width=0.5\textwidth]{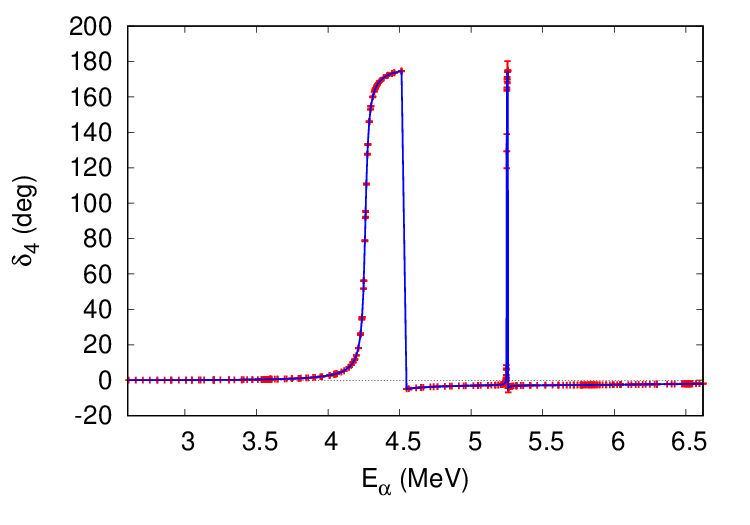}
  \includegraphics[width=0.5\textwidth]{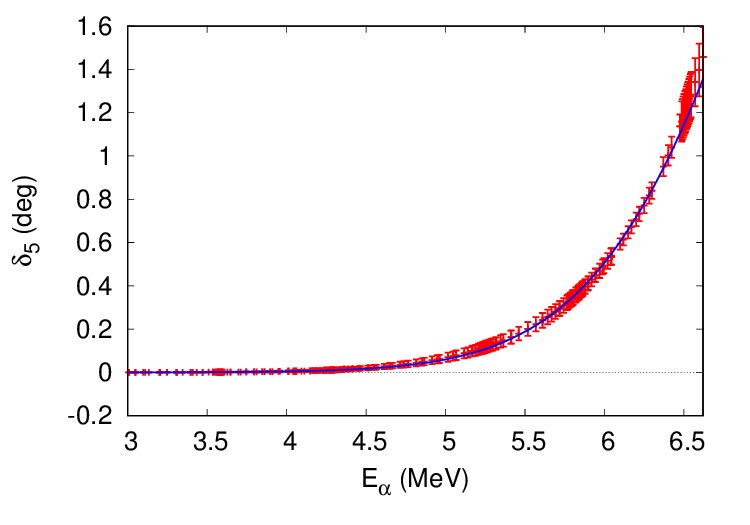}
  \includegraphics[width=0.5\textwidth]{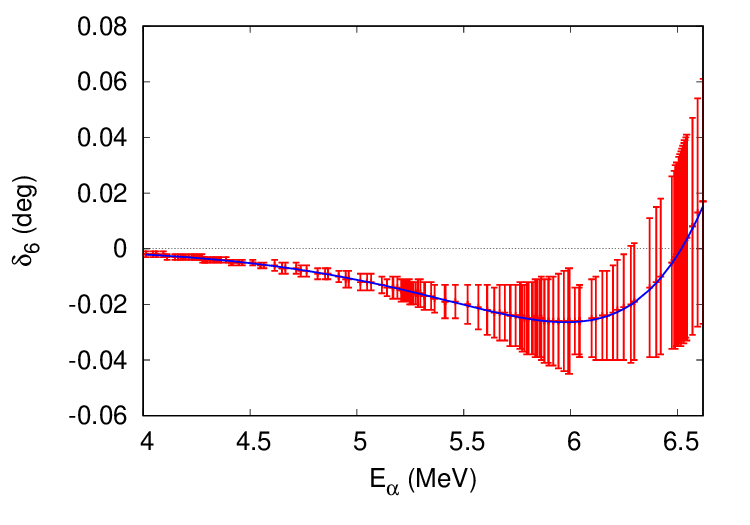}
\caption{
Phase shifts of elastic $\alpha$-$^{12}$C scattering for 
$l=0,1,2,3,4,5,6$
plotted by using the fitted parameters. The experimental data are 
included in the figures as well. 
}
\label{fig;del0123}       
\end{figure}

The parameters are fitted to the accurate phase shift data reported
by Tischhauser et al.~\cite{tetal-prc09}. The fitted values of parameters
can be found in Table II in Ref.~\cite{sa-prc23} (and in Appendix B in 
Ref.~\cite{sa-23} for the inclusion of ground $0_1^+$ state of $^{16}$O). 
Using the fitted values of parameters, we plot the curves of phase shifts
for $l=0,1,2,3,4,5,6$ in the figures in Fig.~\ref{fig;del0123}. 
The experimental phase shift data are included in the figures as well.
One can see that the fitted curves reproduce the experimental data 
pretty well. 

\section{ANCs of bound states of $^{16}$O}
\label{sec:2}

As seen above, the precise phase shift data are well-fitted by
using the expression of $S$ matrices obtained in Eq.~(\ref{eq;exp2idel_l}).
Those fitted parameters are utilized when we calculate the $S$ factors of
$E1$ and $E2$ transitions of $^{12}$C($\alpha$,$\gamma$)$^{16}$O reaction 
and estimate the $S$ factors at the Gamow-peak energy, $E_G = 0.3$~MeV.
We obtained the ANCs of bound states of $^{16}$O, equivalently, the 
wave function normalization factors, which are calculated by employing the 
formula in Eq.~(\ref{eq;Cb}).  
While some values of the ANCs were reported by using 
other theoretical frameworks as well as other nuclear reactions. 
Thus, it may be worth reviewing the present situations
regarding the ANCs of all the bound states,
$0_1^+$, $0_2^+$, $1_1^-$, $2_1^+$, $3_1^-$ states of $^{16}$O.   

\begin{table}
\caption{
ANCs, $|C_b|_1$ and $|C_b|_2$, 
of $1_1^-$ and $2_1^+$ states of $^{16}$O in the literature
and present work
}
\label{table:ANC_12}       
\begin{tabular}{l | ll | l}
\hline 
 & $|C_b|_1$ $(\textrm{fm}^{-1/2})$ & $|C_b|_{2}$ $(\textrm{fm}^{-1/2})$ & 
Reaction \\
\hline 
Brune (1999)~\cite{betal-prl99} & 
 $2.08(20)\times 10^{14}$ & $1.14(10)\times 10^5$ & 
 $^{12}$C(${}^{6(7)}$Li,$d(t)$)${}^{16}$O \\ 
Belhout (2007)~\cite{betal-npa07} & 
 $1.87(54)\times 10^{14}$ & $1.40(50)\times 10^5$ & 
 ${}^{12}$C(${}^6$Li,$d$)${}^{16}$O \\
Oulebsir (2012)~\cite{oetal-prc12} & 
 $2.00(35)\times 10^{14}$ & $1.44(28)\times 10^5$ & 
 ${}^{12}$C(${}^7$Li,$t$)${}^{16}$O \\
Avila (2015)~\cite{aetal-prl15} & 
 $2.09(14)\times 10^{14}$ & $1.22(7)\times 10^5$ & 
 ${}^6$Li(${}^{12}$C,$d$)${}^{16}$O \\
Adhikari (2017)~\cite{aetal-jpg17} & 
 $2.55(35)\times 10^{14}$ & $1.67(23)\times 10^5$ & 
 ${}^{12}$C(${}^7$Li,$t$)${}^{16}$O \\
Orlov (2017)~\cite{oin-prc17} & 
 $2.073 \times 10^{14}$   & $0.5050 \times 10^5$ & 
 ${}^{12}$C($\alpha$,$\alpha$)${}^{12}$C \\
This work~\cite{sa-prc23}       & 
 $1.727(3)\times 10^{14}$ & $0.31(6) \times 10^5$ &
 ${}^{12}$C($\alpha$,$\alpha$)${}^{12}$C \\
\hline
\end{tabular}
\end{table}

The ANCs of the sub-threshold $1_1^-$ and $2_1^+$ states of $^{16}$O have
been intensively studies because the $S$ factors of $E1$ and $E2$ transitions
of $^{12}$C($\alpha$,$\gamma$)$^{16}$O reaction are enhanced because of 
those states.
One can find the summary for the ANCs of $1_1^-$ and $2_1^+$
states of $^{16}$O in Table XIII in deBoer et al.'s paper~\cite{detal-rmp17}.
In Table \ref{table:ANC_12}, we present the recent results from the $\alpha$
transfer reactions and other theoretical estimates 
by Orlov, Irgaziev, and Nabi~\cite{oin-prc17}. 
One can see that the values of ANCs, $|C_b|_1$ and $|C_b|_2$, of 
$1_1^-$ and $2_1^+$ states of $^{16}$O are converging except for those of the 
$2_1^+$ states deduced from the elastic $\alpha$-$^{12}$C scattering data. 

One may notice that our value of ANC of $1_1^-$ in the table,
$1.727(3)\times 10^{14}~\textrm{fm}^{-1/2}$, has a small error bar 
and is remarkably smaller than the others. 
It is still about one $\sigma$ deviation from the values in the table
except for the accurate data of Avila et al.~\cite{aetal-prl15}, 
$2.09(14)\times 10^{14}~\textrm{fm}^{-1/2}$, 
which exhibits about 2.5\,$\sigma$ deviation.  
Recently, a modification of the ANCs of bound states of $^{16}$O
deduced from the $\alpha$-transfer reaction involving $^6$Li 
nucleus was discussed by Hebborn et al.~\cite{hetal-23}. 
They employed a new value of ANC of ground $1_1^+$ state of $^6$Li 
as a two-body $d$-$\alpha$ system, which is obtained in the 
\textit{ab initio} calculation of $\alpha$($d$,$\gamma$)$^6$Li capture
rate and elastic $d$-$\alpha$ scattering 
at energies below 3~MeV~\cite{hetal-prl22},
and re-estimated the ANCs of sub-threshold bound states of $^{16}$O. 
The multiplication of two squared ANCs of 
$^6$Li ($d$-$\alpha$ cluster system) and $^{16}$O ($\alpha$-$^{12}$C cluster
system) is proportional to the cross section of $\alpha$-transfer reaction,
$^6$Li + $^{12}$C $\to d + {}^{16}$O, in the DWBA analysis. 
Thus, when the value of ANC of $^6$Li is increased, that of 
ANC of $^{16}$O should be decreased.
The reported new value of ANC
of $1_1^-$ state of $^{16}$O is $1.84(9) \times 10^{14}~\textrm{fm}^{-1/2}$,
and agrees with our result within about one $\sigma$ deviation.   

Regarding the discrepancy of ANC of $2_1^+$ state obtained from 
the elastic scattering data, it is well known that a typical value 
of the ANC of $2_1^+$ state obtained by using the effective range
parameters in elastic scattering was 
$0.2 \times 10^5~\textrm{fm}^{-1/2}$~\cite{sa-prc18}; about 6 or 7 times
smaller than those obtained from the $\alpha$-transfer reactions.
While a large value of the ANC of $2_1^+$ state, 
$1.384\times 10^5~\textrm{fm}^{-1/2}$, was reported by Sparenberg, Capel,
and Baye by employing the effective range expansion~\cite{scb-jpcs11}.  
We also obtained a large value of ANC of $2_1^+$ state by imposing 
a condition, 
$\left.dD_2/dp^2\right|_{p=p_{i+1}} <  
 \left.dD_2/dp^2\right|_{p=p_{i  }}$ with $p_{i+1} > p_i$ 
at the energy region, $0<E_\alpha<2.6$~MeV, where the experimental 
phase shift data do not exist
(the condition is that the slope of the inverse of propagator, 
$D_2(p)$, is continuously decreasing), but it has a large error bar, 
$(1.3 \pm 3.0)\times 10^{5}~\textrm{fm}^{-1/2}$ 
(see Table II in Ref.~\cite{sa-prc22} as well).  
To obtain such a large value of the ANC of $2_1^+$ state, 
the inverse of propagator $D_2(p)$ should become vanishingly
small at the small energy region, where 
the center values of $D_2(p)$ become 
smaller than its error bars in the denominator. 
This leads to the large error bar of the ANC of $2_1^+$ state.
Thus, it would be unlikely,
by performing the $\chi^2$ fit, 
to obtain a large value of the ANC of $2_1^+$ state 
fitting the effective range parameters to the data.

\begin{table}
\caption{
ANC, $|C_b|_{01}$, of $0_1^+$ ground state of $^{16}$O in the literature
and present work
}
\label{table:ANC_01}       
\begin{tabular}{l | l | l}
\hline 
 & $|C_b|_{01}$ $(\textrm{fm}^{-1/2})$ & Reaction/Method \\
\hline 
Adhikari (2009)~\cite{ab-plb09} & $13.9\pm 2.4$ & 
 ${}^{16}$O + Pb breakup \\
Morais (2011)~\cite{ml-npa11} & 3390 (WS1) & 
 ${}^{12}$C(${}^{16}$O,${}^{12}$C)${}^{16}$O \\
 & 1230 (WS2) & \\
 & 750 (FP)   & \\
Sayre (2012)~\cite{setal-prl12} & 709 & $R$ matrix \\
deBoer (2017)~\cite{detal-rmp17} & 54 & $R$ matrix  \\
Adhikari (2017)~\cite{aetal-jpg17} & $637\pm 86$ & 
 $^{12}$C($^7$Li,$t$)$^{16}$O \\
Shen (2020)~\cite{setal-prl20} & $337\pm 45$ & 
 $^{12}$C($^{11}$Be,$^7$Li)$^{16}$O \\
Orlov (2017)~\cite{oin-prc17} & 21.76 & 
 ${}^{12}$C($\alpha$,$\alpha$)${}^{12}$C \\
This work~\cite{sa-23}       & 
 $45.5\pm 0.3$ & ${}^{12}$C($\alpha$,$\alpha$)${}^{12}$C \\
\hline
\end{tabular}
\end{table}

In Table \ref{table:ANC_01}, values of the ANC of ground $0_1^+$ state
of $^{16}$O are exhibited. As seen in the table, the reported values are
broadly scattered from 13.9~$\textrm{fm}^{-1/2}$ 
to 3390~$\textrm{fm}^{-1/2}$.
The ANC of ground $0_1^+$ state of $^{16}$O would be important to 
calculated the $E1$ and $E2$ transitions of $^{12}$C($\alpha$,$\gamma$)$^{16}$O 
reaction because it determines the overall factor of reaction amplitudes;
it plays a role of the coupling constant of $\alpha CO$ vertex function
in the final state.  
While the concept of ANC for the ground state of $^{16}$O may be questionable.
The ANC of $0_1^+$ state as the two-body $\alpha$-$^{12}$C cluster system 
in terms of the radial part of the wave function is conventionally defined as
\bea
u_0(r) = |C_b|_{01} W_{-\kappa/\gamma_{01},\frac12}(2\gamma_{01}r)\,,
\eea
at the outside of potential range, $r>R$, where $W_{\alpha,\beta}(z)$ is 
the Whittaker function and $R$ is a range of nuclear interaction. 
A typical length scale between $\alpha$ and $^{12}$C cluster in the ground
state is $r=1/\sqrt{2\mu B_{01}} \sim 1$~fm while the size of $^{16}$O is 
larger than that: the radius of $^{16}$O is
$r_A = r_0A^{1/3}\sim 3$~fm.~\footnote{
The length scales for the other bound states of $^{16}$O read
$r=2.5, 2.6, 5.3, 12~\textrm{fm}$ for the $0_2^+$, $3_1^-$, $2_1^+$, $1_1^-$ 
states, respectively. 
Those for the $0_2^+$ and $3_1^-$ states are marginal, and those
for the $2_1^+$ and $1_1^-$ states indicate a significant separation 
of the $\alpha$ and $^{12}$C clusters. 
} 
Because the ground state of $^{16}$O is well described in terms of 
a closed-shell configuration of 
one-particle states in the shell model calculations, 
it is not clear how one can consistently interpret those two pictures 
for the ground state of $^{16}$O;
one is the two-body bound state of $\alpha$ and $^{12}$C clusters 
where they are located so closely inside of the nucleus, 
and the other is 
the closed shell configuration of single-particle states.

\begin{table}
\caption{
ANC, $|C_b|_{02}$, of $0_2^+$ state of $^{16}$O in the literature
and present work
}
\label{table:ANC_02}       
\begin{tabular}{l | l | l}
\hline 
 & $|C_b|_{02}$ $(\textrm{fm}^{-1/2})$ & Reaction/Method \\
\hline 
Sch\"urmann (2011)~\cite{setal-plb11} & $40^{+270}_{-40}$ & $R$ matrix \\
deBoer (2013)~\cite{detal-prc13} & 1800 & $R$ matrix \\ 
Avila (2015)~\cite{aetal-prl15} & $1560\pm 100$ & 
 ${}^6$Li(${}^{12}$C,$d$)${}^{16}$O \\
Orlov (2017)~\cite{oin-prc17} & 405.7 & 
 ${}^{12}$C($\alpha$,$\alpha$)${}^{12}$C \\
Blokhintsev (2022)~\cite{bkms-epja22} & 
 886 -- 1139 & ${}^{12}$C($\alpha$,$\alpha$)${}^{12}$C \\
This work       & $621\pm 9$~\cite{sa-23} & 
 ${}^{12}$C($\alpha$,$\alpha$)${}^{12}$C \\
 & $370\pm 25$ (w/o $0_1^+$)~\cite{sa-prc23} & 
 \\
\hline
\end{tabular}
\end{table}

In Table \ref{table:ANC_02}, the values of ANC of sub-threshold $0_2^+$ state
of $^{16}$O are exhibited. One can see that the reported values of the ANC
are still scattered from 40 to 1800~$\textrm{fm}^{-1/2}$. The ANC of $0_2^+$ 
state of $^{16}$O would play a minor role in the estimate of $S$ factor of
$^{12}$C($\alpha$,$\gamma$)$^{16}$O reaction at $E_G=0.3$~MeV because it 
appears in the cascade transitions; it gives a small portion of the 
contribution to the total $S$ factor. 
While one may notice that the estimate of the ANC of $0_2^+$ state of $^{16}$O
from the elastic $\alpha$-$^{12}$C scattering data significantly depend on 
the theoretical methods. In the four rows from the bottom of the table, the
same phase shift data of the elastic scattering but the different theoretical 
methods are employed. The sensitivity may stem from the fact that a small
value of the inverse of propagator, $D_0(p)$, appears when it approaches 
to the bound state pole, $D_0(i\gamma_{02})=0$, as we have encountered 
in the case of of the ANC of $2_1^+$ state of $^{16}$O above.   

\begin{table}
\caption{
ANC, $|C_b|_{3}$, of $3_1^-$ state of $^{16}$O in the literature and
present work
}
\label{table:ANC_3}       
\begin{tabular}{l | l | l}
\hline 
 & $|C_b|_{3}$ $(\textrm{fm}^{-1/2})$ & Reaction/Method \\
\hline 
Azuma (1994)~\cite{aetal-prc94} & 121 -- 225 & $R$ matrix \\
Tang (2010)~\cite{tetal-prc10}  & 191 -- 258 & $R$ matrix \\
deBoer (2013)~\cite{detal-prc13} & 150 & $R$ matrix \\ 
Avila (2015)~\cite{aetal-prl15} & $139\pm 9$ & 
 ${}^6$Li(${}^{12}$C,$d$)${}^{16}$O \\
Adhikari (2017)~\cite{aetal-jpg17} & $206\pm 23$ & 
 $^{12}$C($^7$Li,$t$)$^{16}$O \\
This work~\cite{sa-prc23} & 
 $113\pm 8$ & ${}^{12}$C($\alpha$,$\alpha$)${}^{12}$C \\
\hline
\end{tabular}
\end{table}

Finally, in Table \ref{table:ANC_3}, the values of the ANC of sub-threshold
$3_1^-$ state of $^{16}$O are exhibited; one can see that the reported values
are relatively converged in the range, 113 -- 258~$\textrm{fm}^{-1/2}$.
The ANC of $3_1^-$ state of $^{16}$O was important and well studied to 
determine a background contribution from the $f$-wave part 
of $\alpha$-$^{12}$C final state 
in the spectrum of $\beta$-delayed $\alpha$ emission 
from $^{16}$N~\cite{aetal-prc94,tetal-prc10}.
As discussed in the case of ANC of $1_1^-$ state of $^{16}$O above, 
the new value of ANC of $1^+$ state of $^{6}$Li leads to the reduced 
value of 
ANC of $1_1^-$ state of $^{16}$O in the DWBA analysis of $\alpha$ 
transfer reaction. For the value of Avila (2015) in the table, 
$139\pm 9~\textrm{fm}^{-1/2}$, it is re-estimated by using the new 
value of the ANC of $1^+$ state of $^6$Li 
as $122\pm 6~\textrm{fm}^{-1/2}$. One can find a good
agreement with our result within the error bars.

\section{Summary}
\label{sec:3}

In the present work, we reviewed the construction of $S$ matrices of
elastic $\alpha$-$^{12}$C scattering at low energy within the framework
of cluster EFT. The parameters in the $S$ matrices are fitted to the 
accurate phase shift data, and the ANCs of $0_1^+$, $0_2^+$, $1_1^-$,
$2_1^+$, $3_1^-$ bound states of $^{16}$O, as the two-body $\alpha$-$^{12}$C
system, are calculated by using the wave function normalization factors
of $^{16}$O propagators for $l=0,1,2,3$. The ANCs we obtained from the 
elastic scattering data are compared with the other results in the 
literature and we discussed the issues of the ANCs of bound states of 
$^{16}$O. 

%
%
%

%

\begin{acknowledgements}
This work was supported by the National Research Foundation of Korea 
(NRF) grant funded by the Korean government (MIST) (No. 2019R1F1A1040362
and 2022R1F1A1070060). 
\end{acknowledgements}



\end{document}